\documentclass[%
12pt,
superscriptaddress,
showpacs,
showkeys,
 amsmath,amssymb,
 aps,
prl,
floatfix,
]{revtex4-1}

\usepackage{graphicx}
\usepackage{dcolumn}
\usepackage{bm}
\usepackage{hyperref}

\usepackage[version=4]{mhchem}
\usepackage{chngcntr}

\begin{document}

\preprint{APS/123-QED}

\title{Effects of confinement on the dynamics and\\
correlation scales in active fluids}

\author{Yi Fan}
\affiliation{Center for Fluid Mechanics, School of Engineering, Brown University, 184 Hope Street, Providence RI 02912 U.S.A.}

\author{Kun-Ta Wu}
\affiliation{Department of Physics, Worcester Polytechnic Institute, 100 Institute Road, Worcester, MA 01609, U.S.A.}

\author{S.Ali Aghvami}
\affiliation{School of Physics, Brandeis University, 415 South Street, Waltham, MA 02453, U.S.A.}

\author{Seth Fraden}
\affiliation{School of Physics, Brandeis University, 415 South Street, Waltham, MA 02453, U.S.A.}

\author{Kenneth S. Breuer}
 \email{kenneth\_breuer@brown.edu}
\affiliation{Center for Fluid Mechanics, School of Engineering, Brown University, 184 Hope Street, Providence RI 02912 U.S.A.}

\date{\today}

\begin{abstract}
We study the influence of solid boundaries on dynamics and structure of active fluids as the height of the container, $z$, changes. Along the varying dimension, the geometry systematically increases, therefore, the confinement ($z$) transits from ``strong confinement", to ``intermediate confinement" and to ``weak confinement'' (close to "unconfined"). In horizontal dimensions ($x,y$), the system remains ``unconfined''. Through tracking the tracers dispersed in the active fluids in three dimensions we observed that activity level, characterized by velocity fluctuations of flow tracers, increases  as system size increases. Concomitantly, the velocity-velocity temporal correlation changes from weak correlation to strong positive correlation, indicating ``memory'' in active flows. We estimate the characteristic size of the flow structure by integrating the velocity-velocity spatial correlation function. The integral increases as confinement becomes weaker and saturates at approximately 400 microns as the system becomes ``unconfined''. This saturation indicates an intrinsic length scale which, along with the small-scale isotropy, demonstrates the multi-scale nature of this kinesin-driven bundled microtubule system. 
\end{abstract}

\pacs{Valid PACS appear here}
\maketitle

\section{Introduction}
Active fluids, and more specifically, cytoskeletal networks of actin filaments or microtubules, exhibit large-scale non-equilibrium dynamics driven by mesoscopic active stresses, which are generated by nanoscale motion of molecular motors ~\cite{vale1985identification, gelles1988tracking, howard1989movement, urrutia1991purified, hunt1994force, schnitzer1997kinesin}. Boundaries can dramatically transform the dynamics of such cytoskeletal active networks generating coherent flow patterns or self-organized states including, for example vortical, spiral, swirling and other patterns ~\cite{ndlec1997self,pinot2009effects, schaller2010polar, sumino2012large, tee2015cellular, miyazaki2015cell,opathalage2019self}. From a biological perspective, the formation of these patterns underlie biological phenomena such as spindle formation or cytoplasmic streaming~\cite{sawin1992mitotic,wittmann2001spindle,theurkauf1994premature,ueda2010myosin,brugues2014physical}, while from a more physical perspective, active fluids provide a novel system that converts chemical energy into mechanical work at the molecular scale and then generates mesoscopic activity at much larger scales, mediated by the short and long-range interactions intrinsic to the specific active medium.

A particularly versatile model system of active fluids is comprised of $\sim$100 micron long bundles of short microtubule filament, each approximately 1.5 microns long. These bundles extend, buckle and fray, driven by the motion of kinesin molecular motors~\cite{sanchez2012spontaneous}. Most of the studies of the isotropic active phase using this long-lasting active fluid have focused on two different configurations: two-dimensional unconfined thin film and three-dimensional annular geometries. The thin film systems are unconfined in the horizontal plane ($x-y$) and measure $100$ microns or less in the vertical ($z$) axis. The vertical confinement is achieved either using an oil film or a solid boundary. This geometry has been used to characterize the dynamics and structure of the system for a variety of different microscopic constituents ~\cite{henkin2014tunable, chandrakar2018microtubule}. 

The second broad category of geometries are three-dimensional annular ``race-tracks'', which have a planform, or $x-y$ geometry, in the form of a circle, square, ratchet, toroid or long channel, but are confined in the transverse, or $y-z$ plane, by solid boundaries. Such confinement can transform the chaotic dynamics of 3D bulk isotropic fluids into long-ranged and long-lived coherent flows, and capable of transporting materials over macroscopic scales~\cite{wu2017transition,varghese2020confinement}. The emergence of such coherent flows appears to be determined by a scale-invariant criterion that depends on the aspect ratio of the confining geometry. 

To understand these effects, it is essential to elucidate the influence of the hard wall boundaries on the active fluid structure and dynamics, and to quantify the changing length and time scales that emerge in response to confinement. In this work we report on the behavior of a 3D bulk active fluid system as it is subjected to decreasing confinement in one direction. To study this, we have conducted a series of experiments in cuboid channels whose horizontal dimension is large ($2$ mm or $4$ mm) compared to any internal scale of the active medium, but whose vertical dimension, $H$,  ranges from a highly confined geometry ($H = 100$ $\mu$m) to one whose scale is large and comparable to the horizontal system scale ($H >2$ mm). In each of these test geometries we tracked passive tracer particles in all three dimensions and calculated their transport, spatial and temporal correlation in the active fluids.
   
\section{Materials and Methods}

\subsection{Active fluids}
Active fluids comprised of microtubule filaments, kinesin-streptavidin motor clusters, depletion agents (Pluronic F-127, Sigma, P2443) and an adenosine triphosphate (ATP) regeneration system.
Pluronic micelles force the microtubule filaments to form bundles \cite{hilitski2015measuring}. Kinesin-streptavidin motor clusters simultaneously bind to and walk along neighbor microtubule filaments, inducing a sliding motor force between anti-polar filaments. The active stress generated by thousands of molecular motors drives the system out of equilibrium. The ATP regeneration system maintains the ATP concentration in the active fluid which ensures that the kinesin motors step at constant speed for the duration of the experiment. . We dispersed $\sim$0.001$\%$ $(v/v)$ spherical colloidal fluorescent tracers ($2.9$ $\mu$m diameter, $395$ nm/$428$ nm, Bangs Laboratories, or $1$ $\mu$m diameter, $412$ nm/$447$ nm, Thermal Fisher Scientific) in the  active fluid, with the tracer size determined by the channel size and target observation region.  Complete details on the active material preparation are described in Supplementary Materials.

\subsection{Test enclosure}
Test chambers, consisting of a rectangular cavity measuring 2 mm by 2 mm ($x-y$), with a height ($z$) that varied from $H = 100$ $\mu$m to $2$ mm (Table~\ref{tab:geometry} (iii-viii)) and a cube measuring $4$ mm in each of the three dimensions (Table~\ref{tab:geometry} (ix)) were fabricated from cyclic olefin copolymer (COC) sheets using a CNC mill (MDA Precision LLC., Model V8-TC8 3-axis). 
The active fluids were loaded into the chamber which was then sealed using a glass slide (Fisher scientific, 12-542C). To ensure a consistent surface condition, all channels and cover slides were coated with polyacrylamide before experiments \cite{sanchez2012spontaneous, henkin2014tunable} (see Supplementary Materials). The assembly was held onto the microscope stage by a machined aluminum plate with a polydimethylsiloxane (PDMS) layer ($\sim$1 mm thickness placed between the glass and aluminum to provide stress relief). 

\subsection{Measurement procedures}
An epi-fluorescence microscope (Nikon TE 200 eclipse) equipped with a $z$-scanning system (Piezo z-drive, Physik Instrumente, Model P-725) was used to observe the motion of the fluorescent tracers.  Light from a high power LED source ($405$ nm, Thorlabs, DC4100) was directed through a narrow-band filter to an air immersion objective (either Nikon CFI S Plan Fluor ELWD $20\times$, NA = $0.45$, WD = $6.9$ mm or Nikon CFI Plan Apo VC $20\times$, NA = $0.75$, WD = $1.0$ mm). The emission from the fluorescent tracers was directed through a narrow-band emission filter and recorded using an sCMOS camera (PCO-Tech, PCO edge 5.5, $2560 \times 2160$ pixels) at $10$ Hz.

Tracer particles in the focal plane appear as sharp spots in the camera image while particles located away from the focal plane appear as ``Airy rings'' \cite{Born-Wolf} whose diameter can be accurately correlated with the particle position above or below the focal plane \cite{afik2015robust,taute2015high} (Fig~\ref{fig:Expt_Methods}A,B).  A typical image contained 30 particles, each of which was tracked in three dimensions using its position in the image and the size of the tracer's point spread function (PSF). The ring size was determined using a Sobel edge detection algorithm and the Circle Hough Transform~\cite{kanopoulos1988design,ballard1981generalizing} (Fig.~\ref{fig:Expt_Methods}C,D). A reference library, which relates the Airy ring size to the tracer's distance from the focal plane, was built (Fig.~\ref{fig:Expt_Methods}E) by $z$-scanning a $1$ $\mu$m tracer particle immobilized in agarose gel with $0.1$ $\mu$m intervals using the piezo z-drive. The resultant measurement and tracking system could determine the tracer location within a $350$ $\mu$m $\times$ $350$ $\mu$m $\times$ $\sim$40 $\mu$m observation volume, with an accuracy of $0.2$ $\mu$m in the $x-y$ plane and $0.4$ $\mu$m in the $z$-direction.

Tracer particle trajectories were assembled from a sequence of images using the ``nearest neighbor'' method \cite{adrian2011particle} and velocities in all three directions ($u_i = u_x, u_y, u_z$) were calculated using finite differences. Following the turbulence literature \cite{tennekes1972first}, velocities were decomposed into mean and fluctuating components: $u_i(t) = \overline{u}_i + u'_i(t)$, where $\overline{u}_i$ denotes an average over all particles tracked during a period of constant system activity ($t = 2000 - 5600$ seconds, Fig.\ref{fig:lifetime}(a)). The variance of the velocity, $\overline{u_i'^2} \equiv \sigma_i^2 =   <(u_i(t)-\overline{u}_i)^2>$, was computed by averaging over all particles tracked during the same time period.

\section{Results and Discussion}

\subsection{Validation of measurement system}
To validate the three-dimensional measurement system, we recorded the Brownian motion of fluorescent tracers ($1$ $\mu$m diameter or $2.9$ $\mu$m diameter) diffusing in deionized (DI) water in three different COC chambers: $H = 4$ mm, $H = 2$ mm and $H = 0.1$ mm. The out-of-focus tracers were tracked from frame to frame in three dimensions based on their Airy ring sizes. As expected, the MSD increased linearly with time (Fig~\ref{fig:validation}) with normalized slopes of $1.03$, $1.00$ and $1.07$ for diffusion in the $2$ mm, $100$ $\mu$m and $4$ mm test enclosures respectively. The deviation from a slope of one could be due to uncertainty in the tracer particle size or a discrepancy between the fluid temperature in the test chambers and the system temperature sensor, located nearby. The measurements were taken in the mid-plane of the chamber, at least $50$ $\mu$m from the nearest solid boundary (in the case of the $100$ $\mu$m enclosure). For a Newtonian fluid, this is sufficiently far from walls such that there should be no confinement effects \cite{happel2012low, lin2000direct, huang2007direct}; as expected, none were observed.

\subsection{Active fluid measurements}
\begin{figure}
\includegraphics[width=5in]{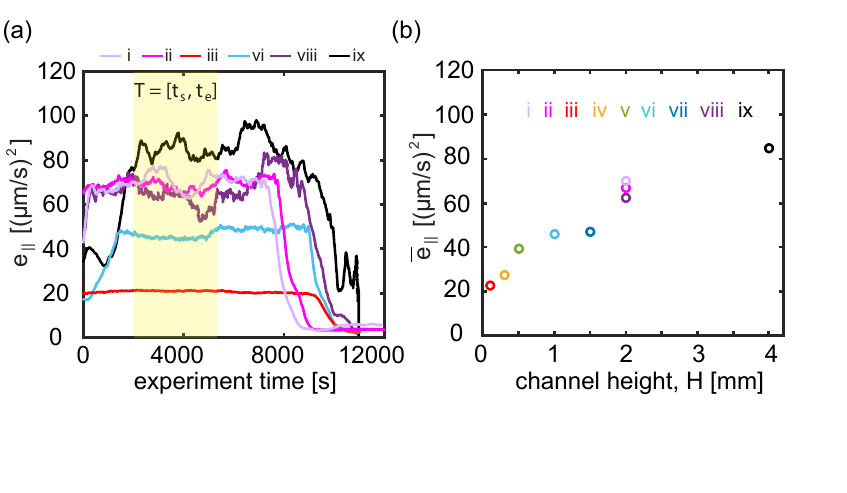}
\caption{\label{fig:lifetime} Color online. (a) Lifetime of active systems in varying geometries, indicated by the parallel component of the specific kinetic energy $e_{\parallel}$. The highlighted area defines a time period, $T = [t_s, t_e]$, over which temporal averaging was calculated. (b) Mean value of parallel component of the specific kinetic energy $\overline{e}_{\parallel}$. The details of roman numerals (i-ix) in the plots are explained in Table~\ref{tab:geometry}.}
\end{figure}

To quantify the system dynamics we measured kinetic energy per unit mass, which was divided into components parallel to ($x-y$: $\parallel$) and perpendicular to ($z$: $\perp$) the nearest confining boundary:
\begin{equation}
e(t) = \frac{E(t)}{\frac{1}{2}m} =  e_\parallel(t) + e_\perp(t) = \left[u_x^2(t) + u_y^2(t)\right] + u_z^2(t) .
\label{eq:kinetic}
\end{equation}
Since we were not able to measure $e_{\perp}$ for the largest channel due to the optical limitations, we compare $e_{\parallel}$ for systems with different levels of confinement (Fig.~\ref{fig:lifetime}a), and observe a consistent behavior: the system evolves quickly to a relatively stable state, remains at that state for approximately 8,000 - 10,000 seconds and then decays. Having the same amount of chemical fuel and the same concentration of molecular motors, all the samples exhibited comparable lifetimes independent of confinement size. This suggests that the energy consumption rate per unit mass is independent of the system size. 

All quantities reported hereafter are computed during the steady-state period between $t = 2000 - 5600$ (Fig.~\ref{fig:lifetime}(a), yellow band).  During this time the average value of the kinetic energy $e_{\parallel}$  increases linearly with the sample thickness, starting at about 20 $(\mu$m/s$)^2$ and reaching a saturation around 80 $(\mu$m/s$)^2$ (Fig.~\ref{fig:lifetime}(b)). 

\begin{figure}
\includegraphics[width=5in]{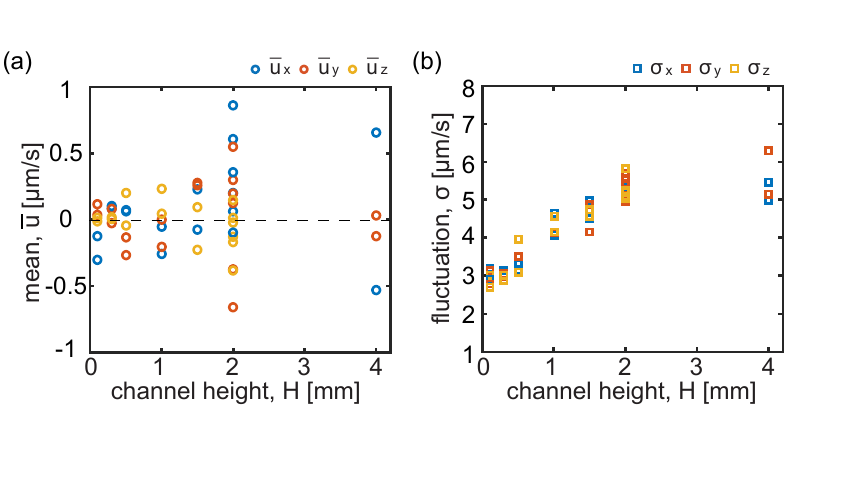}
\caption{\label{fig:velocity} Color online. (a) Mean velocity $\overline{u}_i$ of all tracked particles over the period of $[t_s, t_e]$. Blue: $\overline{u}_x$, red: $\overline{u}_y$, yellow: $\overline{u}_z$. (b) Velocity fluctuation strength $\sigma_i$ of all tracked particles over the period of $[t_s, t_e]$. Blue: $\sigma_x$, red: $\sigma_y$, yellow: $\sigma_z$. Duplicate symbols represent two repeated tests in identical geometries.}
\end{figure}

The time-averaged mean velocities in the $x, y$ and $z$ directions exhibit no long-time sustained large-scale coherent flows (Fig.~\ref{fig:velocity}(a)), in contrast to those previously observed in droplets or ``race-track'' geometries \cite{suzuki2017spatial, wu2017transition}. In addition, the individual components of the velocity fluctuations, $\sigma_i$, are isotropic (fig~\ref{fig:velocity}(b)), exhibiting no preferred orientation, reinforcing the idea that these fluctuations are generated at the smallest scales and that the filaments are, on average, randomly oriented. This is in contrast to the studies of active fluids that were highly confined using an oil surface or a hard wall ~\cite{sanchez2012spontaneous, opathalage2019self}, or systems that exhibit local nematic order near walls ~\cite{wu2017transition} or systems that are liquid crystals in nematic phase \cite{hitt1990microtubule, sanchez2012spontaneous}.

\begin{figure}
\includegraphics[width=5in]{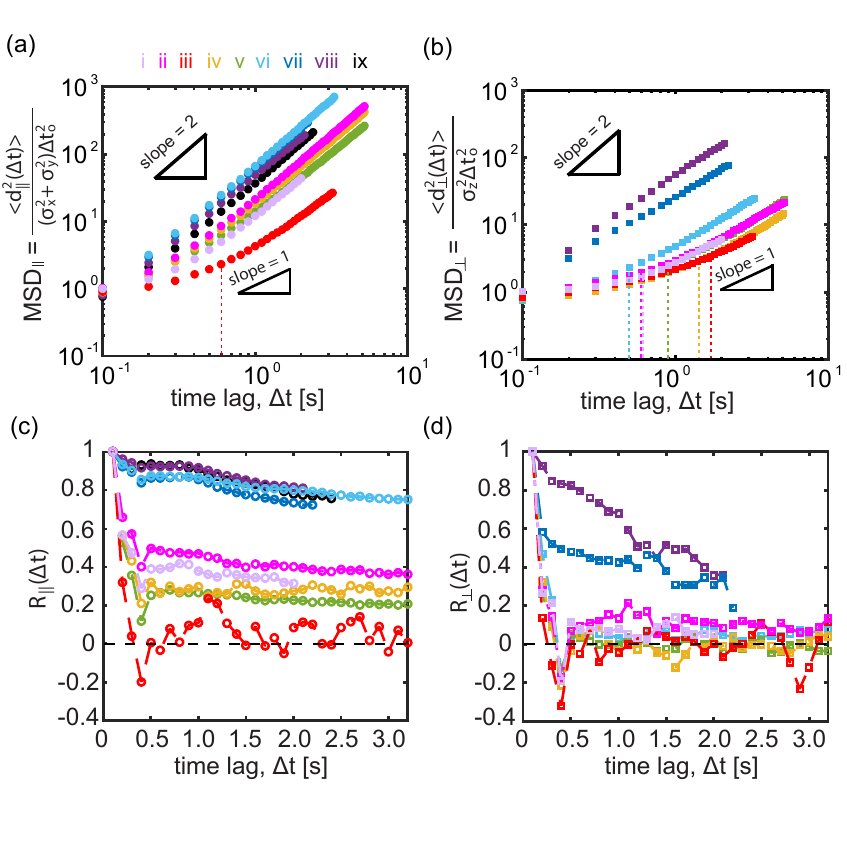}
\caption{\label{fig:Lagrangian} Color online. (a)-(b) Scaled mean square displacements of tracers in the $x-y$ plane ($MSD_{\parallel}$) and in the $z-$direction ($MSD_{\perp}$) in different geometries. Two repeated experiments from the same geometry are averaged together. Dashed lines indicate the crossover time as slope equals 1. (c)-(d) Velocity-velocity temporal correlation in $x-y$ plane and in $z-$direction from different geometries.  The details of roman numerals (i-ix) in the plots are explained in Table~\ref{tab:geometry}.
}
\end{figure}

We compute the scaled mean square displacements (MSDs) of tracers in $x-y$ plane and $z-$ axis:
\begin{subequations}
\begin{eqnarray}
 MSD_\parallel &=& \frac{< (x(t + \Delta t) - x(t))^2 + (y(t + \Delta t) - y(t))^2 >}{(\sigma_x^2+\sigma_y^2)\Delta t_o^2}, \label{eq:scaled_MSDa}
 \\
 MSD_\perp  &=& \frac{(z(t + \Delta t) - z(t))^2 }{\sigma_z^2\Delta t_o^2}. \label{eq:scaled_MSDb}
\end{eqnarray}
\end{subequations}
The normalization accounts for different levels of activity at each confinement.  
The mean square displacement curves exhibit multiple regimes transitioning from sub-diffusive or diffusive to super-diffusive (including ballistic) behavior with increasing time lag $\Delta t$ (Fig.~\ref{fig:Lagrangian}a, b). The dashed lines indicates cross over time lag where the slope of scaled MSD equals $1$. In the parallel MSD, sub-diffusion is observed only in the most confined system, transitioning to a diffusive behavior at $\Delta t \sim 0.6$ second. In contrast, the wall-normal MSD, transitions from sub-diffusive to diffusive are observed for wider range of confinements, with the sub-diffusive regime present in all systems with $H \le 1 $mm, and with the crossover time increasing as the confinement increases. This sub-diffusive MSD may be associated with the frustration of particle motion at the smallest scales by the microtubule network, which becomes increasingly compressed in the wall-normal direction by the confining boundaries. However, this remains to be explored further.

Velocity-velocity correlations,  defined parallel and perpendicular to the confining walls were also calculated:
\begin{subequations}
\label{eq:temporal}
\begin{eqnarray}
 R_{\parallel}(\Delta t) &=& \frac{
  <\mathbf{u}_{\parallel}(t) \cdot \mathbf{u}_{\parallel}(t+\Delta t)>}
 {<\mathbf{u}_{\parallel}(t) \cdot \mathbf{u}_{\parallel}(t+\Delta t_o)>}, \label{eq:temporala}
 \\
 R_{\perp}(\Delta t) &=& \frac{
 <\mathbf{u}_{\perp}(t) \cdot \mathbf{u}_{\perp}(t+\Delta t)>}
{<\mathbf{u}_{\perp}(t) \cdot \mathbf{u}_{\perp}(t+\Delta t_o)>}. \label{eq:temporalb}
\end{eqnarray}
\end{subequations}
Here, $\mathbf{u}_{\parallel}(t)= (u_x(t), u_y(t), 0)$ is the horizontal velocity vector,  and $\mathbf{u}_{\perp}(t) = (0,0,u_z(t))$ is the vertical velocity component.  The trends that were observed in the MSDs are also observed in the velocity-velocity (Lagrangian) self-correlations of flow tracers [Fig.~\ref{fig:Lagrangian}(c, d)]. In strong confinement, the temporal correlations drop almost immediately to zero, indicating diffusive behavior in which flow tracers quickly lose any memory of their motion, in agreement with the behavior observed in the diffusion-dominated MSDs and suggests a lack of any large-scale advective structures in strong confinment. The differences between the perpendicular and parallel correlations $R_\perp, R_\parallel$, echo the above observation that superdiffusion is suppressed in the direction normal to the confining walls. For taller test chambers (H $\geq$ 1 mm), $R_{\parallel}$ exhibits a long-lasting positive correlation while $R_{\perp}$ changes gradually from a flow with a strong long-time coherence to one with almost no correlation (Fig.~\ref{fig:Lagrangian}c, d). Both of these trends agree with the observed ballistic motion and super-diffusion in the scaled MSDs. The long lasting temporal correlations suggest the presence of coherent eddies in the system that continue with a primary direction and magnitude for multiple seconds, and which dominate over the random small-scale uncorrelated motion associated with individual filament motion that characterizes the flow under strong confinement. The absence of average velocities or long-time advective transport implies that in these geometries, unlike the cylindrical or ``race-track'' geometries \cite{wu2017transition},  any coherent eddies or ``rivers'' are not sustained for a long time, or if they are, they re-orient frequently so that the correlation eventually decays. However, the finite field of view of the measurement does not allow for long-enough tracking of a tracer that is required to capture this re-orientation and/or eddy breakup.   

The  coherent structure of the flow can be assessed using the spatial, particle-particle, correlation in the horizontal plane: 
\begin{equation}
\label{eq:spatial}
C_{\parallel}(\Delta r) = \frac{<\mathbf{u}_{\parallel}(\mathbf{r}, t) \cdot \mathbf{u}_{\parallel}(\mathbf{r}+\Delta r, t)>} {<\mathbf{u}_{\parallel}(\mathbf{r}, t) \cdot \mathbf{u}_{\parallel}(\mathbf{r}+\Delta r_o, t)>}, 
\end{equation}
where  $\mathbf{u}_{\parallel}(\mathbf{r},t) = (u_x(\mathbf{r},t), u_y(\mathbf{r},t), 0)$ is the parallel velocity director, $\Delta r$ is the horizontal separation $\Delta r = \sqrt{\Delta x^2 + \Delta y^2}$ and the averaging is performed over all equal-time particle pairs in the time window $[t_s,t_e]$. The correlation is normalized by the value of $C_\parallel$ at a fixed separation $\Delta r_o$ ($\sim30$ $\mu$m). Since particle tracking is limited to a $\sim$40 $\mu$m slab in the $z$-direction, neither a meaningful perpendicular component of the spatial correlation $C_{\perp}$ nor a fully three-dimensional spatial correlation can be calculated.

\begin{figure}
\includegraphics[width=5in]{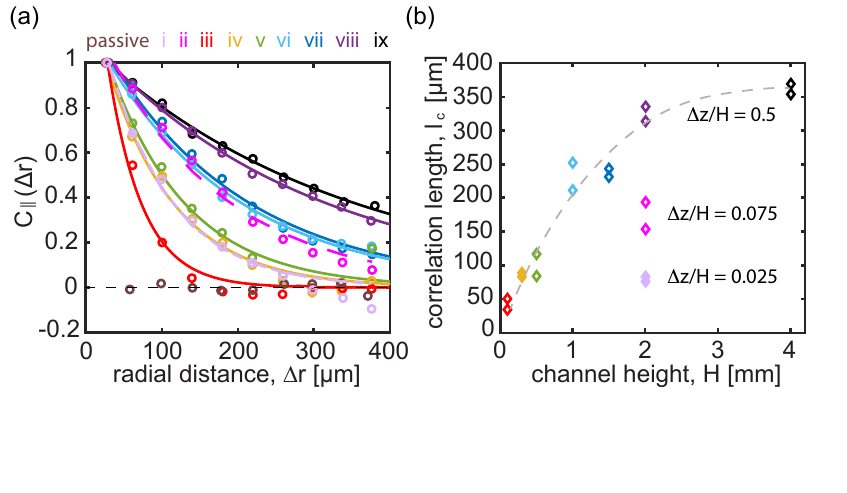}
\caption{\label{fig:spatial} Color online. (a) Normalized equal-time spatial velocity-velocity correlation as a function of separation distance for varying confinements. Circles denote average spatial correlation function measured from two duplicate experiments in each geometry and solid lines present exponential fitting. Brown circles specify the correlation in an ATP-depleted (``dead'') system measured in a 2 mm cubic channel. Solid lines indicate exponential fit to the  data.  (b) Estimated correlation length as a function of channel height computed from the exponential fits. Duplicate symbols represent two repeated tests in identical geometries. The grey dotted line is solely to guide the eye for the measurements made at the center plane of the channel, $\Delta z/H \sim 0.5$  The details of roman numerals (i-ix) in the plots are explained in Table~\ref{tab:geometry}.}
\end{figure}

The correlation function of an ATP-depleted (``dead'') system in an $H = 2$ mm cube drops immediately in a  passive system (Fig.~\ref{fig:spatial}(a) - brown circles). In comparison, active systems exhibit an extended correlation that decays to zero, but whose characteristic length scale depends on the system size (Fig.~\ref{fig:spatial}(a)). To quantify this, we fit an exponential function to the data: 
\begin{equation}
C_\parallel(\Delta r) \approx Ae^{b\Delta r}, 
\label{eq:fitting}
\end{equation}
and extrapolate the fitted function as $\Delta r_o \rightarrow \infty$. Integrating the curve yields a correlation length, $l_c = -A/b \cdot e^{b\Delta r_o}$. As the channel height increases [cases (iii)-(ix)], the correlation length $l_c$ initially grows rapidly, appearing to asymptote to $\sim$400 $\mu$m for largest confinements considered. The measured correlation lengths were close to half of the confinement scale for smallest geometries, $H$, while for the largest confinements they were less than one tenth of the system size. For the $H = 2$ mm cubic confinement, these dynamics were measured at three different distances, $\Delta z$, from the wall [cases (i), (ii) and (viii)]. The correlation length $l_c$ increased as the observation location moved away from the wall and towards the geometric center of the test chamber. Cases (i) and (iii) were observed at the same distance, $\Delta z$, from the boundary/boundaries, however Case (i) was confined by one wall (single-wall) while Case (iii) by two walls (double-wall). A similar comparison exists between cases (ii) and (iv). Comparing the single-wall and double-wall confinement with the same $\Delta z$ [cases (i) and (iii), (ii) and (iv)], reveals that the two-wall confinements exhibited even smaller correlation lengths. 

Long-ranged correlation lengths demonstrate the existence of confinement-size-dependent characteristic eddies in the active flow that are both much larger than the $\sim 1 \mu$m constituent microtubule filament length and considerably smaller than the shortest distance to a confining surface. Comparing with other systems using kinesin motors confined to $50\sim100$ $\mu$m height geometries, similar correlation functions were observed~\cite{henkin2014tunable,chandrakar2018microtubule,lemma2019statistical}. For the largest sized chambers, the correlation length, $l_c$ saturates at $\sim$400 $\mu$m suggesting the existence of an maximum intrinsic  length scale for this active system. Although the correlation length is consistent with the existence of the ``vortex-like'' flow structures that were described by \citet{sanchez2012spontaneous} and \citet{henkin2014tunable}, the shape of the spatial correlation [Fig.~\ref{fig:spatial}(a)] seems to be at odds with this view, as a vortex structure should also result in a negative correlation at some distance, reflecting the return flow of the coherent structure. A possible explanation for this is that the structures we see here are randomly-oriented 3D vortices and the correlation, $C_{\parallel}$, averages over all possible horizontal projections, thus the anti-correlated flow is washed out and not visible in the statistics. In the largely unconfined systems, these vortical structures are long-lived until they dissipate and re-form (or simply rotate) with a different orientation. Thus we observe highly-correlated temporal correlations [Fig.~\ref{fig:Lagrangian}(c)-(d)], super-diffusive and ballistic scaled MSDs [Fig.~\ref{fig:Lagrangian}(a)-(b)], high-level specific kinetic energy [Fig.~\ref{fig:lifetime}(a)] and stronger activity fluctuations [Fig.~\ref{fig:lifetime}(b)-(c)]. Conversely, in the highly confined geometries, the large vortical structures cannot sustain themselves and are replaced by smaller, randomly-oriented, flow structures with short correlation lengths [Fig.~\ref{fig:spatial}(b)] and short lifetimes. In this limit we observe a fast decay in the spatial correlation [Fig.~\ref{fig:spatial}(a)], a rapid drop in the temporal correlation [Fig.~\ref{fig:Lagrangian}(c)-(d)], sub-diffusive and diffusive scaled MSDs for short time lags [Fig.~\ref{fig:Lagrangian}(a)-(b)], lower specific kinetic energy [Fig.~\ref{fig:lifetime}(a)] and weaker activity fluctuation [Fig.~\ref{fig:lifetime}(b)].

\section{Conclusions}
Vortex-like structures have been observed in numerous confined active systems, including cytoskeletal networks, living cells and bacterial suspensions \cite{ndlec1997self, sumino2012large, riedel2005self, dombrowski2004self, dunkel2013fluid}. In the kinesin-microtubule system these long-ranged vortex structures have been observed in highly confined systems \cite{sanchez2012spontaneous}. Here we extend those observations, observing that the correlation scale increases as the wall effects recede.  Although there is the strong suggestion of a saturation in the correlation length (Fig. 4(b)), this is in contrast to recent theories and experiments that suggest the boundaries influence the system at every length scale \cite{PhysRevLett.125.257801} and only experiments in even larger test cells and additional computations will fully resolve this issue.

The results also raise numerous questions that remain to be addressed with further experiments. The present analysis is based on the initial 3600 seconds of the system's activity - around one third of the total system lifetime (more recent kinesin-microtubule systems have demonstrated lifetimes in excess of 70,000 seconds \cite{chandrakar2018microtubulebased}).  A natural question is whether the active fluid has reached a steady state. For instance, if the confinement geometry,  or the constituents of the active fluid are changed, is it possible for stable coherent flows (similar to \cite{wu2017transition}) to emerge, perhaps through some kind of instability \cite{PhysRevLett.125.257801}.  These open questions can be addressed in the future by extending the analysis to cover longer times and in larger systems, as well as observing the structure of the microtubule motion simultaneously with the passive tracer transport.

A second critical unknown is how the surface boundary condition might influence the observed results. Different chemical treatments of the surface affect microtubule adhesion, orientation, order and motion, and will likely affect the system lifetime, as well as the velocity patterns and correlations in the interior, especially in the highly confined geometries. This too must be addressed in follow-up experiments by careful control and variation in the surface chemistry. Numerical simulations (e.g. \cite{varghese2020confinement}) will also be of great value in answering these questions. 
 
\section{acknowledgments}
We thank Dr. Jean Bernard Hishamunda, Dr. Feodor Hilitski and Dr. Stephen DeCamp for the help in protein purification, and Pooja Chandrakar for helpful discussions.  A particular debt of gratitude is due to Zvonimir Dogic for his contributions. YF, KSB and SF designed the experiments. YF, K-TW and SAA prepared the microfluidic devices and samples. YF performed the experiments and analyzed the data. All authors participated in the data analysis, writing and editing the paper. The work was supported by NSF-MRSEC-1420382,  NSF-1336638 and NSF-MRSEC-2011486. YF gratefully acknowledges computation resources from the Brown University Center for Computation \& Visualization (CCV).

\bibliography{confinement}

See Supplemental Material at [URL will be inserted by publisher] for experimental preparations and details.

\newpage
\begin{table*}
\caption{\label{tab:geometry} Experimental details in different channels.}
\begin{ruledtabular}
\begin{tabular}{c|ccc|cc|cc|cc}
 Index & \multicolumn{3}{c|}{chamber dimension} & \multicolumn{2}{c|}{objective} & \multicolumn{2}{c|}{observation position}  & tracer & tracking\\
 number & W & L & H & WD & NA & $\Delta z$ & $\Delta z$/H & diameter  & dimensions\\
  & [mm] &[mm]  &[mm]  &[mm]  & & [mm] & & [$\mu$m] & \\\hline
 i & 2.0 & 2.0 & 2.0 & 1.0 &0.75 & 0.05 & 0.025 & 1.0 & 3D\\
 ii & 2.0 & 2.0 & 2.0 & 1.0 &0.75 & 0.15 & 0.075 & 1.0 & 3D\\
 iii & 2.0 & 2.0 & 0.1 & 1.0 &0.75 & 0.05 & 0.5 & 1.0 & 3D\\
 iv & 2.0 & 2.0 & 0.3 & 1.0 &0.75 & 0.15 & 0.5 & 1.0 & 3D\\
 v & 2.0 & 2.0 & 0.5 & 1.0 &0.75 & 0.25 & 0.5 & 1.0 & 3D\\
 vi & 2.0 & 2.0 & 1.0 & 1.0 &0.75 & 0.50 & 0.5 & 1.0 & 3D\\
 vii & 2.0 & 2.0 & 1.5 & 1.0 &0.75 & 0.75 & 0.5 & 1.0 & 3D\\
 viii & 2.0 & 2.0 & 2.0 & 1.0 &0.75 & 0.80 \footnote{Objective Nikon CFI Plan Apo VC $20\times$. Its working distance (WD) restricted the observation position $\Delta z$.} & 0.4 & 1.0 & 3D\\
ix & 4.0 & 4.0 & 4.0 & 6.9 & 0.45 \footnote{Objective Nikon CFI S Plan Fluor ELWD $20\times$. Its numerical aperture (NA) restricted the tracking dimension to two-dimensional.} & 2.00 & 0.5 & 2.9 & 2D\\
\end{tabular}
\end{ruledtabular}
\end{table*}

\newpage

\clearpage
\newpage
\setcounter{page}{1}
\setcounter{figure}{0}
\setcounter{section}{0}
\renewcommand\thefigure{SI.\arabic{figure}} 
\renewcommand\thesection{SI.\arabic{section}} 

{\centering {\Large \textbf{Supplementary Information}

Effects of confinement on the dynamics and
correlation scales in active fluids\\

Fan et al.}
}

\section{Material preparation details}
Raw tubulin is purified from bovine brains through two cycles of polymerization-depolymerization in high-salt $1$ M $1,4$-piperazinediethanesulphonic (PIPES) buffer at pH $6.8$ and stored at $-80^{\circ}$C \cite{castoldi2003purification}. Before fluorescent labeling or mixing with molecular motors, raw tubulin is recycled for a third time polymerization-depolymerization and flash-frozen by liquid nitrogen at a high concentration (usually larger than $20$ mg/mL, $44.9$ mg/mL in this case) using thin-walled tubes. For fluorescent microscopy, a portion of recycled tubulin at high concentration is labeled with Alexa Fluor 568 (Thermal Fisher Scientific, A20003) by a succinimidyl ester linker. Spectrum absorbance indicated that $34\%$ of tubulin monomers are labeled. To prepare microtubule for the active fluids mixture, recycled tubulins (not fluorescent) are co-polymerized with Alexa 568 labeled tubulins at $37^{\circ}$C for $30$ min to produce microtubules with $3\%$ label percentage. For encouragement of polymerization, $600$ $\mu$M GMPCPP (guanosine-$5'$-[($\alpha$,$\beta$)-methyleno]triphosphate (Jena Biosciences, NU-4056) and $1$ mM dithiothreital (DTT) and M2B buffer is added to target polymerization concentration at $8$ mg/mL. The polymerization concentration can affect polymerization speed and final length of microtubule filaments. The resulting length has an average value of $\sim$1 $\mu$m \cite{sanchez2012spontaneous, wu2017transition}. Microtubules are stored in small aliquots of $10$ $\mu$L at $-80^{\circ}$C. 

We use K401 derived from \textit{Drosophila melanogaster} kinesin, which consists of 401 amino acids of the N-terminal motor domain with a biotin tag \cite{huang1994drosophila}. The biotin tag helps to assemble kinesin motors to multi-motor clusters using streptavidin tetramers (Invitrogen, S-888). Linking kinesins into motor clusters permits binding among multiple microtubule filaments and introduces inter-filament sliding. The mixing ratio of kinesin motors (1.5 $\mu$M) and streptavidin (1.8 $\mu$M) is $1:1.2$ in M2B buffer (M2B: $80$ mM PIPES, $1$ mM EGTA, $2$ mM \ce{MgCl_2}, pH 6.8). $120$ $\mu$M DTT is added before $30$ min incubation at $4^{\circ}$C. After preparation, the kinesin-straptavidin mixture is stored in aliquots at $-80^{\circ}$C. 

Microtubules and motor clusters are prepared separately and combined with other mixtures in a high-salt buffer (M2B + $3.9$ mM \ce{MgCl_2}). We add an ATP regeneration system, as the most important energy supply for motor clusters, using $26$ mM phosphoenolpyruvate (PEP) and $2.8\%$ ($v/v$) stock pyruvate kinase/lactic dehydrogenase enzymes (PK/LDH) (Sigma, P0294). Motor clusters hydrolyze ATP to ADP while stepping on microtubule filaments. PK consumes PEP and converts ADP back to ATP, therefore, the ATP concentration in the active systems remain stable until PEP and ATP are depleted. To reduce photobleaching effect, we minimize the oxygen exposure of fluorescent samples by adding $2$ mM trolox (Sigma, 238813) and anti-oxidants, including $0.22$ mg/mL glucose oxidase(Sigma, G2133), $0.038$ mg/mL catalase (Sigma, C40) and $3.3$ mg/mL glucose. We add $5.5$ mM DTT to stabilize proteins, $2\%$ ($w/w$) Pluronic F127 to force microtubule filaments into bundles and $\sim$0.001\% ($v/v$) fluorescent particles to probe the active systems. The final concentration of microtubules is $1.3$ mg/mL, kinesin is $1.5$ $\mu$M and ATP is $1.4$ mM.

\section{Microchannel channel surface preparation}
To ensure a consistent surface, a standard procedure was followed:
\begin{enumerate}
   \item Immerse and sonicate silicon channels for $5$ minutes in boiled $1\%$ $(v/v)$ Hellmanex in deionized (DI) water. 
    \item Rinse sonicated devices thoroughly in DI water and then soak them in ethanol for $10$ minutes.
    \item Rinse devices thoroughly in DI water and soak them in $1$ M \ce{KOH} solution for $10$ minutes. 
    \item Prepare silane solution: $100$ mL ethanol, $1$ mL acetic acid and $0.5$ mL trimethoxysilyol-propyl methacrylate (\ce{H2C=C(CH3)CO2(CH2)3Si(OCH3)3}). 
    \item Rinse devices thoroughly in DI water and immerse them in silane solution for $10$ to $20$ minutes. 
    \item  Prepare acrylamide solution. We used $20 \%$ acrylamide solution and dilute it to $2 \%$ with DI water. Then  place $2 \%$ acrylamide solution in vacuum chamber to degas for $10$ minutes. 
    \item Take devices out from silane solution, rinse them with DI water and blow dry them with compressed nitrogen (or compressed air). 
    \item Polymerize acrylamide. Take acrylamide solution out from the vacuum chamber. Mix $100$ mL $2 \%$ acrylamide solution with $70$ mg ammonium persulfate (APS) and $35$ $\mu$L tetramethylethylenediamine (TEMED). Note: let the APS dissolve first while stirring and then add TEMED. Volume and mass can be linearly increased if larger volume of polyacrylamide solution in need. 
    \item  Within $20$ seconds after adding TEMED, immerse dry devices into polymerized acrylamide solution. Coating should be ready in $\sim$3 hours. 
\end{enumerate}

\newpage
\begin{figure}
\includegraphics[width=5in]{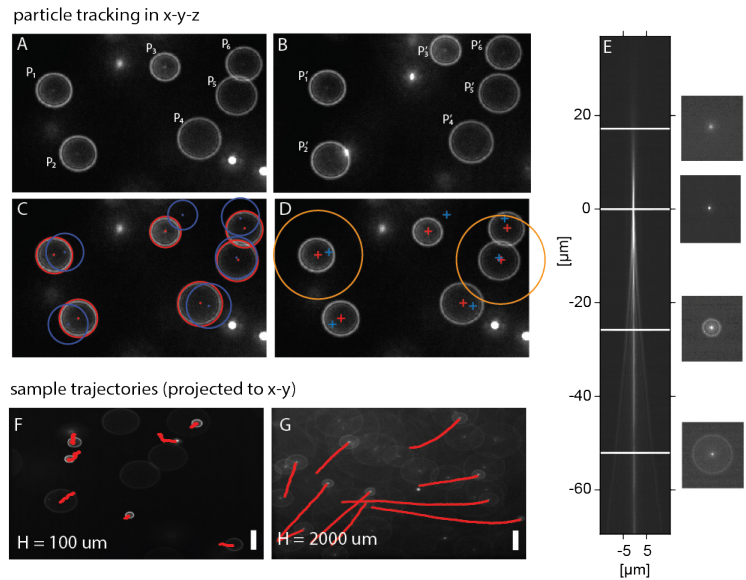}
\caption{\label{fig:Expt_Methods} Example of particle tracking in three dimensions.  A,B: raw images taken at two successive times illustrating the sparse seeding of tracer particles.  Particle positions are indicated by the Airy rings. C: The center and size of each ring is detected using Sobel edge detection and the Hough circle transform.  D: Particles are tracked according to a nearest neighbor algorithm. E: The $z-$location of each particle is localized using the library of ring sizes vs. distance from the focal plane obtained from a calibration. F,G: Two samples of particle tracks, taken from a shallow and a deep test geometry ($H = 100$, $2000$ $\mu$m) illustrating the qualitative change in particle mobility due to confinement.  The scale bar represents 40 $\mu$m.}
\end{figure}

\begin{figure}
\includegraphics[width=3in]{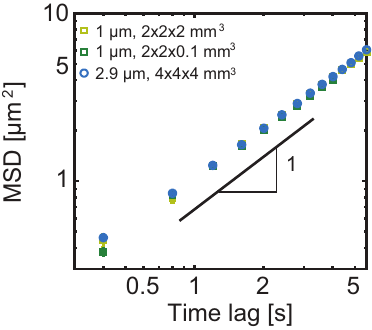}
\caption{Normalized Mean Square Displacements (MSDs) of 1 and 2.9 $\mu$m particles diffusing in water and tracked using the Airy-disk three-dimensional tracking technique. The MSDs 
are normalized by the size of particle ($1$ $\mu$m or $2.9$ $\mu$m diameter), and the error bars indicate standard errors of the averaging over all tracked particles.}
\label{fig:validation}
\end{figure}

\end{document}